\documentclass{aa}
\usepackage{graphicx}
\begin{document}
   \title{IR spectra of the microquasar GRS~1915+105 during a low state}

   \author{E. T. Harlaftis
          \inst{1}
          \and
          V. R. Dhillon\inst{2}
	\and
		 A.~Castro-Tirado\inst{3,4}
          }

   \offprints{E. T. Harlaftis}

   \institute{Institute for Astronomy and Astrophysics, 
			National Observatory of 
Athens, P. O. Box 20048, Athens 118 10, Greece\\
              \email{ehh@astro.noa.gr}
         \and
Department of Physics and Astronomy, University of Sheffield, Sheffield 
S3 7RH, UK\\
             \email{}
	\and
Instituto de Astrof\'{\i}sica de Andaluc\'{\i}a (IAA-CSIC), P.O. Box 03004,
   E-18080 Granada, Spain\\
	\and
	Laboratorio de Astrof\'{\i}sica Espacial y F\'{\i}sica Fundamental 
   (LAEFF-INTA), P.O. Box 50727, E-28080 Madrid, Spain   
          \email{}
             }

   \date{Received January 15, 2001; accepted , 2001}

   \abstract{
There is controversy regarding the nature of  the suspected donor star
to the microquasar GRS~1915+105,  and hence whether GRS~1915+105  is a
high mass X-ray binary  (HMXB) or a low-mass  X-ray binary (LMXB).  In
order  to clarify this issue, we  obtained  an infrared  (J,  H and K)
spectrum of  GRS~1915+105  in its  low   state which shows  a  steeper
continuum than the prototype X-ray binary  Sco~X-1.  We did not detect
any He{\small~II} emission  at 2.189 $\mu$m  from GRS~1915+105  in our
quiescent spectrum, indicating that the  line is transient and is only
observed  during episodes of high  X-ray activity.  For our instrument
configuration, there is no detection of the  $^{12}$CO lines which are
characteristic of late-type stars either in GRS1915+105 or Sco X--1.
   }

\authorrunning{Harlaftis et al.}
\titlerunning{IR spectra of GRS1915+105}
   \maketitle
%
%________________________________________________________________

\section{Introduction}

The galactic X-ray source GRS~1915+105 (Castro-Tirado et al. 1994) was
the first  object in our Galaxy  to show superluminal jets  (there are
about   10  microquasars  in   our  Galaxy    now; see  Mirabel    and
Rodr\'{\i}guez 1999).  Until  then, apparent  superluminal motion  had
only been observed in active galactic nuclei, which  are thought to be
powered by supermassive black holes. The proximity of GRS~1915+105 (12
kpc; Chaty  et al.  1996) and the   short-term variations it exhibits,
provide  an excellent test-bed to  investigate the detailed physics of
relativistic jets. Ejection of  relativistic plasma clouds in the form
of synchrotron flares at infrared and radio wavelengths (Fender et al.
1997; Pooley and Fender 1997; Eikenberry et  al.  1998; Mirabel et al.
1998) results in the rapid disappearance  of the inner accretion disc,
following the model  proposed by Belloni et  al.  (1997), offering the
first convincing connection   between accretion discs and  jets. Also,
GRS~1915+105 is probably the heaviest known stellar-mass black hole in
the Galaxy; the 67 Hz QPO present  in the X-rays  implies a mass of 33
M$_{\odot}$ for a spinning black hole, assuming that the QPO arises in
a Keplerian orbit at the inner accretion disc (Morgan et al. 1997; Cui
et al.  1998).

Due    to the   large  optical  extinction   towards  the  source,  IR
spectroscopy,  coupled with simultaneous  RXTE and radio observations,
is the only possible way of gaining further insight into the nature of
this exotic system.   In particular, IR  spectroscopy  has caused some
debate on the correct model  of the system  based on the visibility of
the He{\small~II} emission line at 2.189  $\mu$m.  The K-band spectrum
of GRS~1915+105 shows   strong   He{\small~I} (2.059 $\mu$m)    and  
Br$\gamma$  (2.166 $\mu$m) emission   lines.  Eikenberry et al.  (1998)
found flux variations of a   factor of 5  in these  lines  on a 5  min
timescale.  During flares, the line fluxes varied linearly with the IR
continuum flux, implying that the  lines are radiatively pumped by the
flares.   He{\small~II}   (2.189 $\mu$m) emission  has   been reported
(Castro-Tirado et al.   1996; Eikenberry et  al. 1998) whereas Mirabel
et al. (1997) did not observe any trace of it.  Based on the detection
or not of  the He{\small~II} emission  line, it was suggested that the
companion could be a LMXB or a HMXB, respectively.  Since the spectral
type of the  companion star of  GRS~1915+105 is still an open question
we attempt,  in  this paper,  to clarify  the issue  by revisiting the
object and comparing the HK spectrum with that of the prototype X-ray 
binary Sco~X-1.
 
%__________________________________________________________________

\section{Observations}

%                                     Two column figure (place early!)
%______________________________________________ Gamma_1 (lg rho, lg e)
   \begin{figure*}
   \centering
   \includegraphics{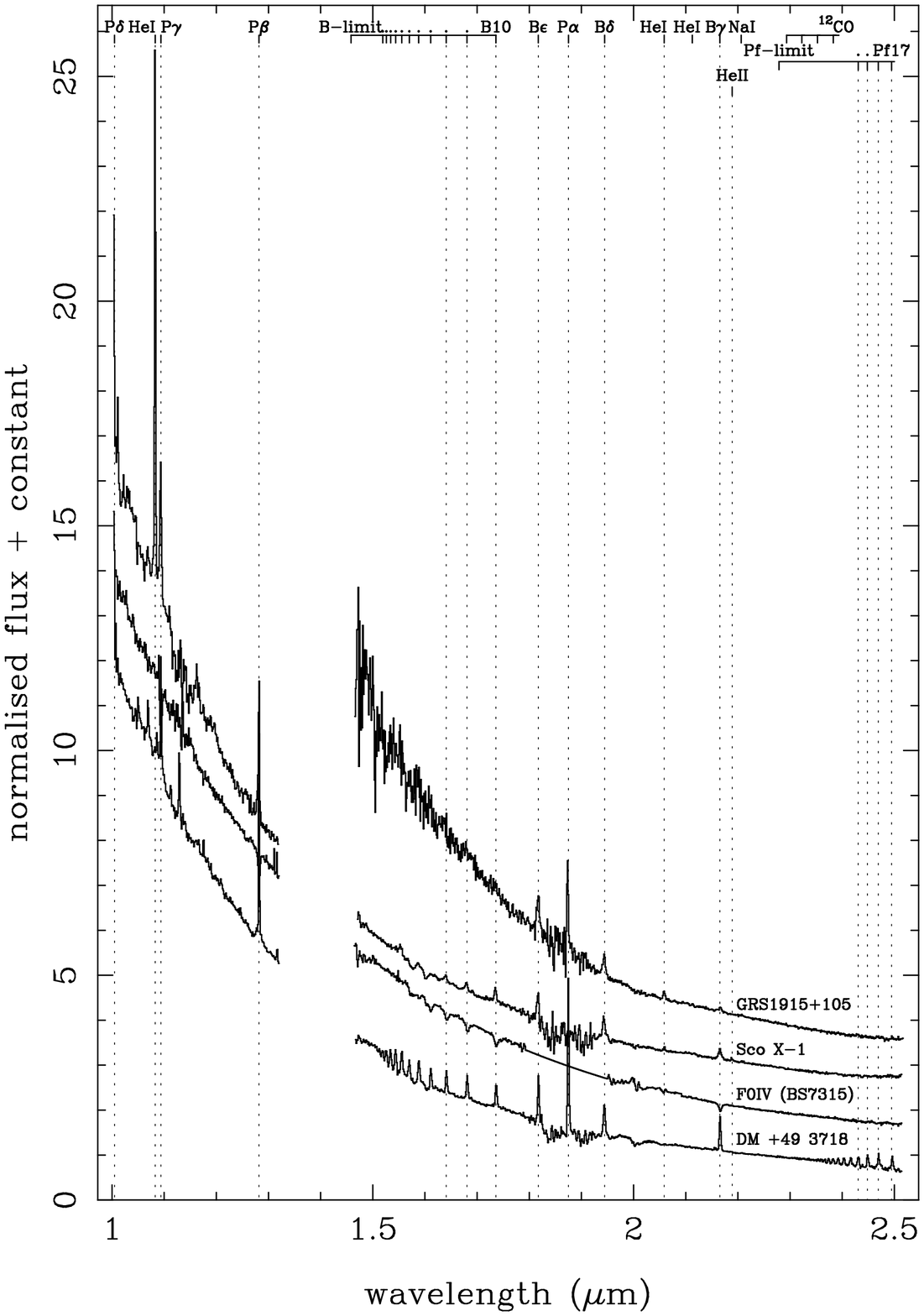}
   \caption{The  deredenned HK spectrum  of the microquasar  GRS~1915+105 and the
JHK  spectra of Sco~X-1 and the  Be star DM +49 3718.  Also plotted is
the  JHK spectrum  of  the F0IV star BS~7315.  The   spectra have been
normalised by dividing by the flux at 2.250  $\mu$m and then offset on
the $y$-axis by adding a multiple of 1 to each spectrum.}
              \label{fig1grs}%
    \end{figure*}

We obtained J (1.002-1.322 $\mu$m), H  (1.455-2.094 $\mu$m) and K-band
(1.906-2.547 $\mu$m) spectra  of GRS~1915+105 using the Cooled Grating
Spectrometer (CGS4) on   the 3.8m  United Kingdom  Infrared  Telescope
(UKIRT)  on  Mauna Kea  during  the night  of 1999  July 8 (journal of
observations  in Table 1).   The 40  lines  mm$^{-1}$ grating was used
with the 150 mm camera and the 256$\times$256 pixel InSb array, giving
resolutions of 324, 426 and 340  km s$^{-1}$ in  the J, H and K-bands,
respectively.  We also obtained JHK spectra of Sco~X-1 and the Be star
DM  +49 3718 in  order  to compare them  with  the spectral slope  and
emission line properties  of  GRS~1915+105.  The target   spectra were
bracketed   by observations of  nearby F-type  stars,  which were used
during the data reduction to remove the telluric atmospheric features.
We utilized the non-destructive  readout mode to minimize the  readout
noise.  The  spectra were obtained    with the one  pixel slit   (1.23
arcseconds) and were sampled over two  pixels by mechanically shifting
the array in 0.5 pixel  steps in the  dispersion direction. We exposed
for a maximum of 2 minutes on the object and then shifted the spectrum
onto a different spatial position on  the array (by nodding the target
along the slit)  in order to  provide accurate  sky subtraction.   The
nodding was  repeated until a good signal  to noise ratio was achieved
for the target spectrum.  The CGS4 data reduction system performed the
initial data reduction steps (application of  the bad pixel mask, dark
subtraction, bias subtraction,  flat-field division, sky  subtraction;
see also Daly \& Beard  1994).  The residual  sky background was  then
removed and the object spectra  were optimally extracted using FIGARO.
Ripples   with  a periodicity  of  two    pixels, resulting  from  the
mechanical shifting of the  array, were visible in the one-dimensional
spectra and were subsequently removed  (using IRFLAT in FIGARO as well
as   private software,  for  comparison).   Wavelength calibration was
performed using an Argon arc in the K-band, a  Xenon arc in the H-band
and a  Krypton arc in the  J-band, giving a  root-mean-square error of
$\sim$ 1 \AA \  with a second order  polynomial fit.  Flux calibration
was  performed using the  F-type  standards.  The target spectra  were
divided  by the  closest  observed F star   spectra (with the  stellar
features interpolated across), and then multiplied  by the flux of the
standard at each wavelength, determined from black body functions with
the same effective temperature  and flux as  the standard.  The H-band
and  K-band spectra overlap in the  region between 1.906--2.094 $\mu$m
and we combined them, after matching the flux scale, by averaging them
using pixel weights which optimized the signal-to-noise ratio.

%__________________________________________________ One column table
   \begin{table*}
      \caption[]{Journal of Observations ($t_{exp}$ is the total time exposure,
$\lambda_{cen}$ is the central wavelength of the band observed and
UTC refers to the time at mid-exposure).}
         \label{table1}
	$$
         \begin{array}{p{0.2\linewidth}lcccccc}
            \hline
            \noalign{\smallskip}
            \hline
 Star & t_{exp}& \lambda_{cen}=2.227\mu m & t_{exp} 
     &  \lambda_{cen}=1.775 \mu m & t_{exp}& \lambda_{cen}=1.162\mu m \\ 
            \noalign{\smallskip}
            \hline
     & min &  UTC & min & UTC  & min & UTC	\\
	    \hline
            \noalign{\smallskip}
Sco X-1   	& 16.0&06:20:00 & 24.0 & 10:09:39   & 16.0 & 05:42:25^{\dagger}\\
GRS~1915+105  	& 64.0&07:56:59 & 80.0 & 11:34:10   & 64.0 & 13:48:27		\\
DM +49 3718	& 10.7&09:01:51 & 5.3  & 09:32:08   &  8.0 & 14:47:13		\\
BS~7315		& 0.4 &09:44:59 & 0.8  & 08:52:34   &  0.4 & 14:34:17		\\
            \noalign{\smallskip}
            \hline
         \end{array}
     $$ 
\begin{list}{}{}
\item[$^{\dagger}$] Observed on 1999 August 14. All other
spectra were obtained on 1999 July 8.
\end{list}
   \end{table*}
%
  
%                                     Two column figure (place early!)
%______________________________________________ Gamma_1 (lg rho, lg e)
   \begin{figure*}
   \centering
   \includegraphics{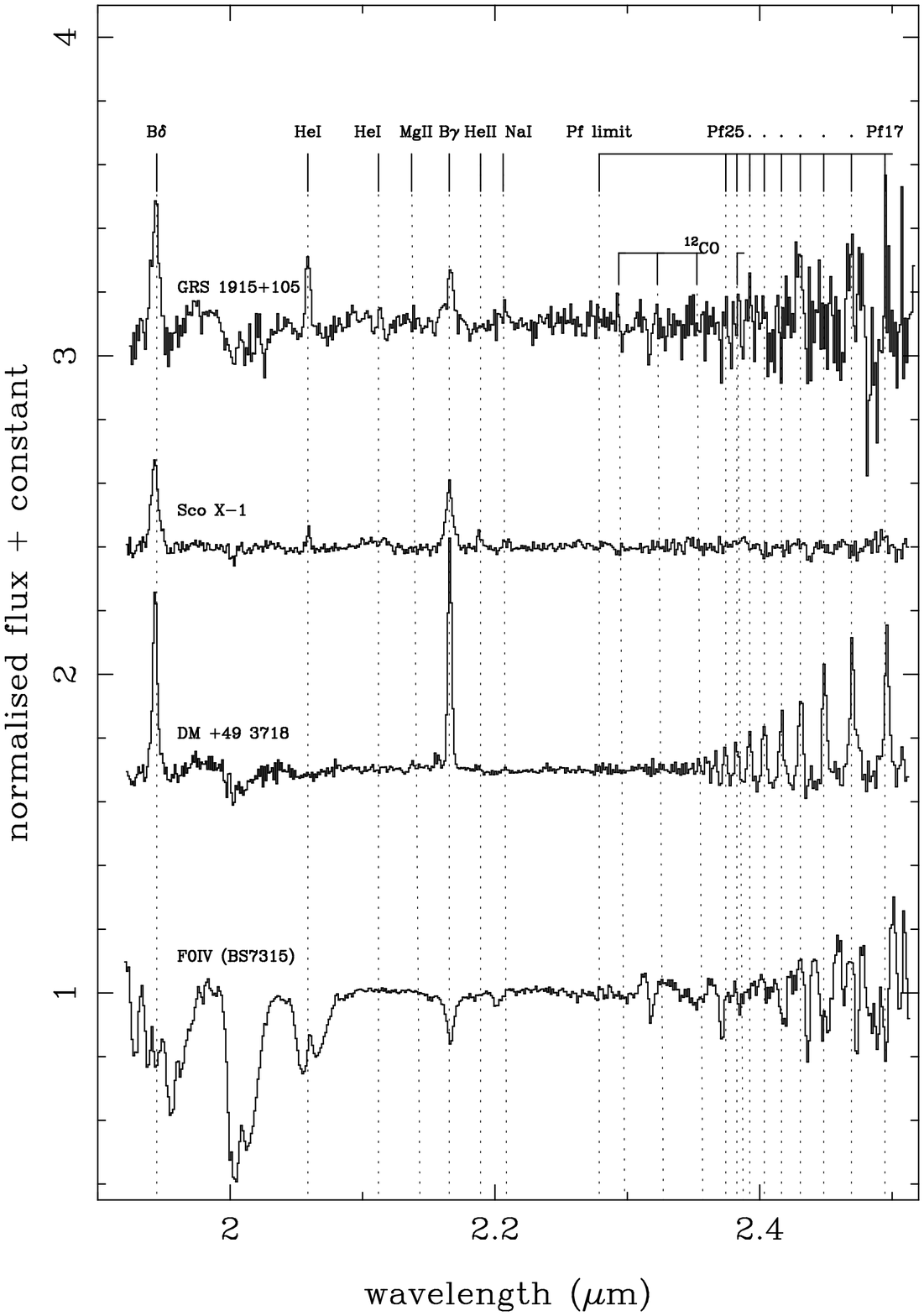}
   \caption{The K-band spectra of the  microquasar GRS~1915+105, Sco X-1, and the
Be star DM +49  3718. Also plotted is  the K-band spectrum of the F0IV
star  BS~7315  which  indicates  the  location of  telluric absorption
features.  The  spectra have been normalized by  dividing  by a spline
fit to their  continua and then   offset on the  $y$-axis  by adding a
multiple of 0.7 to each spectrum.}
              \label{fig2grs}%
    \end{figure*}

\section{A High mass or a Low mass X-ray binary?}

In  Figure 1 we present the  JHK spectra of the Be   star DM +49 3718,
Sco~X--1 and the HK spectrum of  GRS~1915+105 (the J-band spectrum was
not useful due  to  the  large  extinction),  and the   F0IV-type star
BS~7315 for comparison.  For clarity, we show  just the K-band spectra
in Figure 2.   In Table 2,  we list the  wavelengths,  line fluxes and
equivalent widths of the most prominent  lines identified in Figure 1.
Line fluxes have not been corrected for slit losses, hence only ratios
of line  fluxes can reliably be  used.  The line  widths (FWHM) of the
emission lines in 1H~2202+501 and GRS~1915+105 are 610 km s$^{-1}$ and
730 km   s$^{-1}$, respectively.   The  line  width of  Br$\gamma$  in
GRS~1915+105 is {\bf 730  km s$^{-1}$}, consistent with other measured
widths during a low state - since the GRS~1915+105 activity started in
1992 - (760, 730  km s$^{-1}$ measured by Mart\'{\i}  et al. 2000, and
Mirabel et al.   1997, respectively). {\bf  The line width}  increases
during the IR flaring activity (1010  and 1260 km s$^{-1}$ measured by
Mart\'{\i} et al.  2000, and Mirabel et al. 1997, respectively).

\section{The Be star  DM +49 3718 spectrum}

The K-band  spectrum of the Be  star DM +49  3718 was classified  as a
peculiar K-band star, showing Br$\gamma$ in  emission and perhaps some
trace of He{\small~II}  at 2.189 $\mu$m (Hanson  et al.  1996). It has
also been suggested that DM +49 3718 is the optical counterpart of the
High Mass X-ray Binary 1H~2202+501 (Tuohy et al.  1988) but without an
unambiguous confirmation  so  far.  The  equivalent width  (EW) of the
Br$\gamma$ line in our data (see Table 2) is consistent with the value
of 27 \AA  \ measured by  Hanson et al.  (1996).   In our K and H-band
spectra,   the   Pfund   and  Brackett    lines  can  individually  be
distinguished  up to Pf28  and Br20, and  the Pfund and Brackett jumps
are weakly in emission at 2.279 and 1.458 $\mu$m, respectively.  There
are  also  emission lines at 1.069,  1.087  and 1.113 $\mu$m  which we
identify as  Si{\small~ I}.  We identify  the other  emission lines at
1.129 $\mu$m, 1.175 and 1.317  $\mu$m as O{\small~I}, C{\small~I}  and
Ca{\small~I} lines (Melendez and Barbuy 1999).  The power law slope of
the JHK spectrum is not significantly different from that of Sco~X--1,
with a power  law ($F_{\lambda} \propto  \lambda  ^{\alpha}$) index of
$\alpha$=--3.21$\pm$0.01.  The ratios of line intensities of the Pfund
series are consistent   with recombination case  B  (Hummer and Storey
1987)  within the   (large) uncertainties,  indicating  that the  line
source  is located in  optically thin  gas  surrounding the high  mass
star.
%The line ratios can be used to  probe the transition between optically
%thin to  the thick regime.   
%  The Brackett
%series  though  deviate considerably  from  the  recombination B case.
%Since the series jumps  are flat we conclude  that the transition from
%an optically thin gas  to the thick  regime  appears around the  Pfund
%lines 24-25.

\section{The GRS~1915+105 spectrum}

GRS~1905+105 is not a Wolf-Rayet star, like  Cyg X--3 (van Kerkwijk et
al.  1992),  since   we   detect Br$\gamma$  in  emission    (see also
discussion in Castro-Tirado, Geballe and Lund  1996) but no absorption
features  from the companion star (neutral  lines or molecular bands),
as seen    in the   galactic  bulge   sources  GX 1+4    and GX   13+1
(Bandyopadhyay et  al.   1997).  There is   no trace of  He{\small~II}
emission line at a time when the radio spectrum was  flat and thus the
disc  was in a low state.   The last flaring  activity was observed on
1999 July 2.  The radio and X-ray continua were in their lowest states
(Mart\'{\i} et al.  2000) which  indicate that the accretion disc  was
quiet in  terms of its behaviour since  1992.  It  is unclear  if the
He{\small~II} emission is produced along   the jet, in the   accretion
disc or on the massive star.  There are indeed examples of HMXBs, such
as Cyg  X--1, where He{\small~II} emission  is produced in the stellar
wind of  the O9.7 Iab primary star  (Gies and  Bolton 1986).  However,
the X-ray burst activity is related to the  jet/disc system centred on
the compact object (Belloni  et al.  1997),  rather than on a possible
massive companion star,  and thereafter is  the most likely  cause for
pumping up the  transient He{\small~II} emission.  This hypothesis  is
supported by the absence  of He{\small~II} emission during this low state,
which is the time  that the inner  disc is  being replenished by  gas,
suggesting that the X-ray luminosity is  not sufficient to pump up the
He{\small~II} line through photoionization of the accretion disc.

The spectrum of GRS~1915+105 is different from that  of the Be star DM
+49  3718  since there is  no  helium emission in  the  latter and the
hydrogen  lines have narrower  velocity  widths and  larger equivalent
widths.  Although hydrogen and helium emission is  also observed in Be
HMXBs, we   do not detect their    characteristic neutral or molecular
emission lines of Na{\small~I} (2.206  and 2.209 $\mu$m), Mg{\small~I}
(2.281  $\mu$m), and $^{12}$CO bandheads  (2.294,  2.323, 2.353, 2.383
$\mu$m (Everall et al. 1993).  The  latter neutral and molecular lines
are observed in  absorption in LMXBs   and can determine  the spectral
type of the companion  star as in  GX13+1  and G1+4 (Bandyopadhyay  et
al. 1999).  Our K spectrum of GRS~1915+105 does  not show any evidence
of the above trace lines either in absorption or in emission, hindered
by inadequate  signal-to-noise ratio.  A VLT  spectrum on 4 July shows
the trace   lines  of  Na{\small~I}   and  Mg{\small~I}  in   emission
(Mart\'{\i} et al.  2000).  Unfortunately, the wavelength coverage was
short of the $^{12}$CO bandheads.  A higher-resolution VLT spectrum on
20  and  24  July shows  unequivocably   the  $^{12}$CO  bandheads  in
absorption indicating that the companion  star is a G-M  spectral-type
star (Greiner  et al. 2001).   Apparently,  the system has  significant
transient continuum and line emission which either dilutes or fills in
the   late-type's photospheric lines     out of detection  making  the
detection of the late-type star so elusive until now.

We dereddened  the GRS~1915+105  using  E(B--V)=9.6 (Bo\"{e}r  et  al.
1996), A$_{V}$=26.5 (Chaty et  al.  1996) and  the extinction law from
Howarth (1983).   There was insufficient signal  in  the J-band, given
the large  extinction, though  after  smoothing  it was  possible   to
combine the J-band spectrum with the H and K-band  spectra to derive a
power law index of --5.15$\pm$0.02.  The steeper slope in GRS~1915+105
than in the Be star DM +49 3718  indicates that there is a significant
J-band  emission component due to  the accretion disc during this low
state.  For comparison,  the  dominant IR  component is due  to
synchrotron emission  during flares  (Mirabel et al.  1998). 

\section{Conclusions}

Our comparison of the infrared spectra of the microquasar GRS~1915+105
and  the prototype  X-ray binary Sco  X-1  give the following results.
The  spectra of Sco X--1 and  GRS~1915+105 are very similar.  The main
difference between   the Sco  X--1  and  GRS~1915+105  spectra is  the
He{\small~II}  line at 2.189  $\mu$m.  In  Sco X--1, the He{\small~II}
line is persistent  and is most  likely related to irradiation  of the
accretion disc  around   the compact  object.    In GRS~1915+105,  the
He{\small~II}  emission line is  transient and is observed only during
the X-ray  bursting activity  (Mart\'{\i}  et al.  2000).   The  other
difference between the Sco X--1 and GRS~1915+105 K-band spectra is the
weaker  He{\small~I}  line at 2.058  $\mu$m,  which indicates a higher
luminosity class for the donor star  to GRS~1915+105. Finally, we find
a steeper JHK  continuum slope in GRS~1915+105  than in Sco~X--1.  The
detection    of  the $^{12}$CO  bandheads   in  absorption (Greiner et
al. 2001)  is consistent with the  constraints provided  by Eikenberry
and Bandyopadhyay (2001) for  a  Roche-lobe overflow from a  late-type
companion star to GRS~1915+105.  The non-detection of these bands with
our   UKIRT  spectrum may  be   related, {\bf  not  only to  the lower
signal-to-noise    and lower spectral  resolution,  but   also} to the
transient  nature of the  emission lines and  the state the system was
during our observations. {\bf Neutral  spectral features have recently been
detected by Chandra which may be  related to cold material surrounding
GRS~1915+105  (Lee et   al. 2001),  and therefore   the origin of  the
$^{12}$CO bandheads  will  be unequivocably attributed  to  a low mass
star only  after phase-resolved   spectra show  the   expected Doppler
shifts of the bandheads due to the orbital motion.}

%The line ratios   of the  Brackett  series  will be  used  to  extract
%information on   the  electron density   by  comparison to theoretical
%recombination  values, and also will provide  an independent value for
%the  extinction.   
%Additional spectroscopy in the J and H bands  from 8-10m class telescopes
%will allow one to model the  dereddened continuum and distinguish between
%the  different components  of   IR  emission (blackbody from   the
%companion,  free-free   radiation,  reprocessed   emission   from  the
%accretion disc). 

%__________________________________________________ One column table
   \begin{table*} 
  \caption[]{Wavelengths,  equivalent  widths (\AA),\ and line
   fluxes  (in units of   10$^{-14}$,  10$^{-15}$ and  10$^{-13}$ ergs
   cm$^{-1}$ s$^{-1}$ from left to right) of the most prominent lines
   in  the  infrared spectra   of Sco  X-1,   GRS~1915+105, and DM +49
   3718. The last two columns give the  observed line intensity ratios
   and the  theoretical  case  B (for   $N_{e}$ = 10$^{4}$  cm$^{-3}$,
   $T_{e}=10^{4}$ K) line  intensity ratios with respect  to a line in
   the series (Pf18, Br10  and P$\beta$). Note  that the spectra  have
   not been corrected for slit losses and  hence the lines fluxes must
   be used with caution.  The line ratios, however, are secure.}
         \label{table2}
     $$ 
         \begin{array}{p{0.1\linewidth}lcrrcccccc}
            \hline
            \noalign{\smallskip}
            \hline
	 &          &\multicolumn{2}{c}{Sco~ X-1}&\multicolumn{2}{c}{GRS~1915+105}&\multicolumn{2}{c}{DM +49 3718}& &    \\
            \noalign{\smallskip}
            \hline
Line	 & \mu m   &line flux   &	 EW &  line flux& EW      &line flux &	EW  & Int. ratio &Case B \\
            \hline
     \noalign{\smallskip}
P$\delta$& 1.0049   &11.2\pm1.4&  -9\pm2&	    &           		&13\pm2   & -7\pm1   &0.30\pm0.04 & 0.34\\
He~I 	 & 1.0830 &33.3\pm0.7& -26\pm2&	    &           		&   -       & 	-	 &     \\
P$\gamma$& 1.0938   &11.3\pm0.6& -11\pm1&     &           		&11\pm5   & -6\pm1   &0.26\pm0.06 & 0.55\\
P$\beta$ & 1.2818   &10.1\pm0.3& -13\pm1&     &           		&22\pm3   &-23\pm2   &1.00    & 1.00 \\
B20 	 & 1.5196   &          & 	    &     	   &           	&1.0\pm0.3& -3\pm2	 &0.19\pm0.09	&0.13\\
B19 	 & 1.5265   &          & 	    &     	   &           	&1.5\pm0.3& -4\pm2   &0.31\pm0.09	&0.15\\
B18 	 & 1.5346   &          & 	    &     	   &           	&2.2\pm0.3& -5\pm3   &0.46\pm0.09	&0.17\\
B17 	 & 1.5443   &          & 	    &     	   &           	&2.4\pm0.3& -6\pm3   &0.62\pm0.11	&0.20\\
B16 	 & 1.5561   &          & 	    &     	   &           	&3.4\pm0.3& -8\pm3   &0.69\pm0.12	&0.24\\
B15	 & 1.5705   &          & 	    &     	   &           	&2.4\pm0.3& -6\pm3   &0.54\pm0.09	&0.29\\
B14 	 & 1.5885   &          & 	    &     	   &           	&3.4\pm0.3& -9\pm3   &0.73\pm0.11	&0.36\\
B13 	 & 1.6114   & 	       & 	    &     	   &           	&1.9\pm0.2& -5\pm2   &0.65\pm0.10	&0.45\\
B12 	 & 1.6412   &1.0\pm0.2&-3\pm1   &     	   &           	&2.6\pm0.3& -7\pm2   &0.77\pm0.12	&0.57\\
B11	 & 1.6811   &1.4\pm0.2&-4\pm1   &     	   &           	&3.1\pm0.3& -9\pm2   &0.85\pm0.12	&0.75\\
B10 	 & 1.7367   &2.1\pm0.2&-7\pm1   &     	   &           	&4.1\pm0.3&-12\pm2   &1.00		&1.00\\
B$\epsilon$&1.8181  &3.0\pm0.4&-11\pm2  & 	3.1\pm0.4& -18\pm2  &3.3\pm0.3&-24\pm2   &2.34\pm0.25	&1.39\\
P$\alpha$& 1.8751   &          & 	    &     	   &           	&18\pm2   &-72\pm10  &1.65\pm0.44 & 2.05\\
B$\delta$& 1.9451   &4.1\pm0.2&-15\pm2  & 	3.3\pm0.2& -15\pm2  &5.6\pm0.5&-23\pm2   &2.18\pm0.25	&1.99\\
He~I 	 & 2.0587   &0.5\pm0.1 &-2.4\pm1& 	0.9\pm0.2& -9\pm1 	&   -       & 	-	 &     \\
%He~I 	 & 2.1126   &            & 	    &  0.4$\pm$1   &  1$\pm$2 	&           & 		 &     \\
B$\gamma$& 2.1661   &2.2\pm0.1&-15\pm1  &   1.1\pm0.1& -14\pm1  &4.6\pm0.2&-28\pm1   &2.73\pm0.15	&3.03\\
He~II	 & 2.1891   &0.4\pm0.1 &-1.6\pm1&     0	   &     0      &0.03\pm0.08&-0.3\pm1.0&     \\
Pf25 	 & 2.3744   &          & 	    &	     	   &            &0.5\pm0.2& -2\pm2   &0.24\pm0.16 &0.39\\
Pf24 	 & 2.3828   &          & 	    &	     	   &            &0.6\pm0.2& -3\pm2   &0.28\pm0.17 &0.44\\
Pf23 	 & 2.3925   &          & 	    &	     	   &            &0.7\pm0.2& -4\pm2   &0.38\pm0.16 &0.49\\
Pf22 	 & 2.4035   &          & 	    &	     	   &            &0.8\pm0.2& -6\pm2   &0.49\pm0.20 &0.56\\
Pf21 	 & 2.4164   &          & 	    &	     	   &            &1.0\pm0.2& -9\pm2 	 &0.54\pm0.25 &0.64\\
Pf20 	 & 2.4314   &          & 	    &     	   &            &1.4\pm0.3&-11\pm2   &0.74\pm0.26 &0.73\\
Pf19 	 & 2.4490   &          & 	    &     	   &            &1.8\pm0.3&-15\pm2   &1.00\pm0.36 &0.85\\
Pf18 	 & 2.4700   &          & 	    &     	   &            &2.7\pm0.3&-19\pm3   &1.00 &1.00 \\
Pf17	 & 2.4953   &	       &	    &		   &		&2.4\pm0.4&-19\pm3   &1.14\pm0.25	&1.19 \\
            \noalign{\smallskip}
            \hline
         \end{array}
     $$ 

   \end{table*}

\begin{acknowledgements}
      The  data  reduction and  analysis  was  partly  carried  out  at  the
Sheffield STARLINK node.  Use of   software developed by T. R.  Marsh  is
gratefully acknowledged.   The SERVICE programme of  UKIRT was used to
obtain  the  J-band spectrum  of Sco X-1  and we  thank  Tom  Kerr for
observing it.
\end{acknowledgements}

\end{document}